\documentclass[12pt]{iopart}

\usepackage{hyperref}
\hypersetup{colorlinks=true,linkcolor=blue,urlcolor=blue,citecolor=blue}
\usepackage{graphicx}
\usepackage{amssymb}
\usepackage{tikz}
\definecolor{seal}{rgb}{0.26, 0.7, 0.68}
\usepackage{scalerel}
\newcommand*\circled[1]{\tikz[baseline=(char.base)]{
            \node[shape=circle,draw,inner sep=0.5pt] (char) {#1};}}
            \newcommand*\mathbox[1]{\tikz[baseline=(char.base)]{
            \node[shape=rectangle,draw,inner sep=0.5pt] (char) {#1};}}

\begin{document}

\title[]{Features of Magnetically-induced atomic transitions of Rb D$_1$ line studied by Doppler-free method based on the second derivative of the absorption spectra}

\author{Armen Sargsyan$^1$, Arevik Amiryan$^1$,  Emmanuel Klinger$^{2}$ \& David Sarkisyan$^1$ }

\address{$^1$ Institute for Physical Research -- National Academy of Sciences of Armenia, 0203 Ashtarak-2, Armenia}
\address{$^2$ Helmholtz-Institut Mainz -- GSI Helmholtzzentrum f\"ur Schwerionenforschung, Johannes Gutenberg-Universit\"at, 55128 Mainz, Germany}
\ead{eklinger@uni-mainz.de}
\vspace{10pt}
\begin{indented}
\item[]
\end{indented}

\begin{abstract}
We show that the Second Derivative (SD) technique of the absorption spectra of Rb atomic vapours confined in a nanocell with a thickness $\ell= \lambda/2=398$~nm allows to achieve close to Doppler-free spectroscopy. Narrow linewidth and linearity of the SD signal response with respect to transition probabilities allow us to study separately, in an external transverse magnetic field (0.6 to 4~kG), a big number of the atomic transitions of $^{85}$Rb and $^{87}$Rb atoms. Atomic transitions $|F_g,0\rangle \rightarrow |F_e=F_g,0'\rangle$, for which the dipole moment is null at zero magnetic field (so-called magnetically-induced transitions), show a gigantic increase in probability with increasing magnetic field. When a magnetic field is applied on the vapour, we show the possibility of forming a dark resonance on these transitions by adding a coupling laser. We hence demonstrate a five-fold increase in the transmission of the probe radiation when the coupling laser is on. Theoretical calculations are in a very good agreement with experimental results.
\end{abstract}

%
\noindent{\it Keywords}: nanocell, Rb D$_1$ line, Doppler-free spectroscopy, dark resonance, magnetic field

%
%
%

\section{Introduction}

The efforts undertaken in the understanding of magneto-optical phenomena have lead to the development of various applications including narrow-band optical filters based on the Faraday effect \cite{budker.PRA.2002,zentile.JPB.2014,tao.OE.2019}, atomic frequency reference \cite{sargsyan.APL.2008,reed.OSAC.2018}, four-wave mixing processes in strongly shifted atomic levels \cite{whiting.JMO.2018}, wide-band tunable laser frequency stabilization using narrow atomic transitions in strong magnetic fields \cite{sargsyan.OL.2014,mathew.OL.2018}, etc. For theses applications, thermal vapours of alkali metals, in particular rubidium and caesium, are widely used. 

When alkali atoms are placed in an external magnetic field, their energy levels undergo shifts and split into a large number of components, while transitions probabilities experience large changes; therefore precise knowledge of atomic transitions's behaviour in a magnetic field is crucial \cite{auzinsh.Book.2010,olsen.PRA.2011,sargsyan.OL.2012}. In the case of thermal vapours, the motion of atoms leads to the well known Doppler effect, which broadens each Rb lines by about 0.6~GHz at room temperature. In other words, a sub-Doppler resolution is needed to study separately each individual atomic transition between hyperfine splitting (hfs) Zeeman sublevels of the ground and excited states. Indeed, in the case of a natural mixture of $^{85}$Rb and $^{87}$Rb, the number of closely spaced atomic transitions can reach several tens. Sub-Doppler resolution with simple single-beam geometry, providing a linear response of atomic media for transmission, fluorescence, and selective reflection experiments, can be attained using optical nanocells (NCs) with a thickness of the order of resonant radiation wavelength \cite{sarkisyan.OC.2001,sargsyan.JPB.2019}. But even in this case, some spectral features cannot be completely resolved due to residual Doppler-overlapped profiles \cite{das.JPB.2008}. 

It was recently shown \cite{sargsyan.OL.2019a} that a simple data-processing technique allows the retrieval of homogeneous lineshape of individual optical transitions in the absorption spectrum. The implemented technique is especially efficient when the laser is shined on the region where the cell thickness reaches half the laser wavelength as a consequence of the coherent Dicke narrowing \cite{briaudeau.PRA.1998,sargsyan.JPB.2016,amiryan.JMO.2018}. The procedure consists in recording the absorption spectrum $S_a(\nu)$ by scanning the laser frequency across the resonant region, and derive either numerically or with electronics the second-order derivative $S_a''(\nu)$ from it, see \cite{sargsyan.OL.2019a} for more details. As a result, the transition linewidth in the second derivative (SD) spectrum reduces down to 30-40 MHz (FWHM), which corresponds to a $\sim20$-fold narrowing as compared with the Doppler width. The SD technique has been known for a long time \cite{savitzky.AChem.1964,talsky.Book.1994}; yet not really applicable for atomic spectroscopy with ordinary cm-long alkali vapour cells as hyperfine transitions are often completely overlapped.
	
In this paper, we study both experimentally and theoretically the features of Rb D$_1$ line in magnetic fields ranging from 0.6 to 4.0~kG, with a high frequency resolution. We show that frequency positions of spectral features and their amplitudes linked, in the weak-probe limit, with transition probabilities, are preserved with the SD technique.  Such a good sub-Doppler spectral resolution allows us to identify all atomic transitions of the D$_1$ line of a natural mixture of Rb, and carry out quantitative measurements in strong transverse magnetic field, which was not possible in previous works because of insufficiently high spectral resolution \cite{zentile.JPB.2014,weller.JPB.2012,keaveney.JPB.2019}. In these works, isotopically pure vapour cells containing $^{87}$Rb isotope had to be used. Finally, we demonstrate the possibility to form a dark resonance on one of the magnetically-induced transitions (MI) of $^{87}$Rb using a coupling and a probe laser, when a transverse magnetic field is applied on the vapour. In Sec.~\ref{sec:exp}, we detail the experimental setup used to perform our studies. Section~\ref{sec:theory} gives an overview of the theoretical model allowing to calculate both alkali transitions evolution in a magnetic field, and the transition lineshape inherent to the NC. Results are discussed in Sec.~\ref{sec:results} and a conclusion is given in Sec.~\ref{sec:conc}.

\begin{figure}[h!]

\centering
\includegraphics[width=9cm]{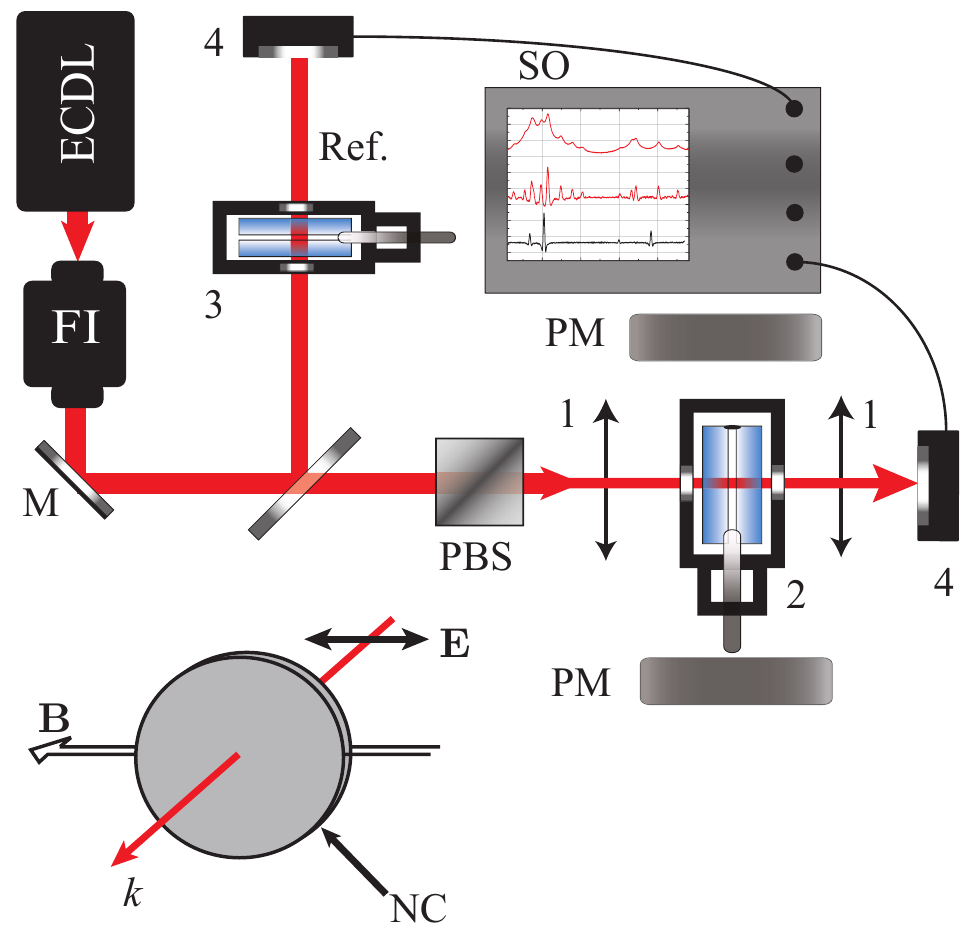}
\caption{Sketch of the experimental setup. ECDL -- extended cavity diode laser, FI -- Faraday isolator, PBS -- polarising beam splitter, PM -- permanent magnets, M -- mirror, 1 -- lenses, 2 -- Rb NC in the oven, 3 -- frequency reference channel, 4 -- photodiodes, SO -- digital storage Tektronix TDS2014B oscilloscope, $\mathbf E$  -- electric field of the laser radiation, $\mathbf B$ -- magnetic field applied to the nanocell such that $\mathbf B \parallel \mathbf E$. Configuration of the magnetic measurement is presented in the lower left inset.}
\label{fig1}
\end{figure}

\section{Experimental arrangement}
\label{sec:exp}

The first design of the T-shaped nanocell, sometimes called extremely thin cell, was presented in \cite{sarkisyan.OC.2001}. It consists of two sapphire windows and a vertical sapphire side arm tube (a metal reservoir). Later, this design has been somewhat modified and a description of NCs' recent  design can be found in \cite{keaveney.PRL.2012}. The NC used in this experiment has 2.4 mm-thick sapphire windows and is filled with a natural mixture of $^{85}$Rb (72.2\%) and $^{87}$Rb (27.8\%); it can operate up to 400$~^\circ$C. To reach sufficient signal-to-noise ratio, the NC is operated with a specially designed oven having two openings for the laser beam transmission. The stem of the NC was heated at 120$~^\circ$C, corresponding to a vapour density $\mathcal N = 2.0\times10^{13}$~ cm$^{-3}$; however the windows were maintained at a temperature that was about 20$~^\circ$C higher to prevent condensation of alkali vapour on NC's windows. The height of the NC in the oven was adjusted such that the laser shines the region $\ell \approx\lambda/2\approx 398$~nm of the NC.

The sketch of the experimental setup is shown in Fig.~\ref{fig1}. A linearly-polarised beam emitted by a VitaWave laser \cite{vassiliev.RSI.2006} (ECDL, $ \lambda = 795$~nm, spectral linewidth $\gamma_\lambda< 2\pi \times 1$~MHz) is tuned in the vicinity of Rb D$_1$ line. To avoid feedback in the cavity, a Faraday insulator is placed right after the laser. A Polarising beam splitter (PBS) is used to purify initial linear radiation polarisation of the laser. A part of the laser radiation was directed to an auxiliary channel composed by a Rb NC (3), from which a Rb D$_1$ line SD spectrum at zero magnetic field is recorded and serves as a frequency reference. The transmission signal was detected by a photodiode (4) and was read by a Tektronix TDS 2014B digital four-channel storage oscilloscope. 

Configuration of the magnetic measurement is presented in the inset of Fig.~\ref{fig1}. The magnetic field $\mathbf B$ is directed along light's electric field $\mathbf E$ (\textit{i.e.} $\mathbf B \parallel \mathbf E$), exciting $\pi$ transitions of the Rb D$_1$ line. Since Helmholtz coils are not suitable for magnetic field over 0.2~kG, we have used two strong permanent magnets (PMs) mounted on a translation stage as a magnetic field control \cite{klinger.AO.2020}. Magnetic strength applied in the vapour was varied by simple longitudinal displacement of the permanent magnets, calibrated with a Teslameter HT201 magnetometer. Note that it was not possible to achieve magnetic fields greater than 4~kG in this configuration due to the transverse size of the oven.

\section{Theoretical considerations}
\label{sec:theory}
In this section, we briefly describe the theoretical model used to simulate the spectra presented in Sec.~\ref{sec:results}. We divide the modelling in two parts: \begin{enumerate}
\item calculation of the evolution of atomic levels and transitions dipole moments as a function of the magnetic field;
\item evaluation of transition lineshape inherent to the NC  and sum over all possible transitions (two-level systems) given a light field polarisation.
\end{enumerate}
The interaction of the vapour with the magnetic field and the optical field is thus treated separately. This simple procedure is in general enough to calculate weak-probe absorption or reflection spectra from NC as was shown by many articles \cite{klinger.AO.2020,sargsyan.JOSAB.2017,sargsyan.JPB.2018,klinger.EPJD.2017}. 
\subsection{Transition frequencies and dipole moment in a magnetic field}
The presence of a static magnetic field modifies both energy levels (shifts) and transition dipole moments as is known for a long time, see e.g. \cite{tremblay.PRA.1990}. The total Hamiltonian $\mathcal H_m$ describing the evolution of atomic levels in a magnetic field is given by the sum of the unperturbed hyperfine levels' Hamiltonian $\mathcal H_{hfs}$ with the one describing atom -- magnetic field interaction; it reads
\begin{equation}
\mathcal H_m = \mathcal H_{hfs} +\frac{\mu_B}{\hbar}B_z(g_LL_z + g_SS_z+g_II_z),
\label{eq:hamil_B}
\end{equation} 
where $\mu_B$ is the Bohr magneton, $L_z$, $S_z$, and $I_z$ are respectively the projection of the orbital, electron spin and nucleus spin momentum along $z$, chosen as the quantization axis; $g_{L,S,I}$ are the associated Land\'e factors. The transition dipole moments $\langle e| d_{q}|g\rangle$ between two states are proportional to the transfer coefficients, such that
\begin{equation}
\langle e| d_{q}|g\rangle \propto \sum_{F_e'F_g'}c_{F_eF_e'}a\left[\Psi(F_e,m_e);\Psi(F_g,m_g);q\right]c_{F_gF_g'},
\end{equation}
with the coefficients $c_{FF'}$ given by the eigenvectors of $\mathcal H_m$, and
\begin{eqnarray}
\fl \eqalign{a\left[\Psi(F_e,m_e);\Psi(F_g,m_g);q\right]=&(-1)^{1+I+J_e+F_e+F_g-m_{Fe}}\sqrt{2J_e+1}\sqrt{2F_e+1}\sqrt{2F_g+1}\\
&\times \left( \begin{array}{r@{\quad}cr} 
F_e & 1 & F_g \\
-m_{F_e} & q & m_{F_g}
\end{array}\right)
\left\{ \begin{array}{r@{\quad}cr} 
F_e & 1 & F_g \\
J_g & I & J_e
\end{array}\right\},}
\end{eqnarray}
where the parentheses and the curly brackets denote the 3-$j$ and 6-$j$ coefficients, respectively; $q=0,\pm1$ is associated the polarisation of the excitation such that $q=0$ for a $\pi$-polarised laser field, $q=\pm1$ for a $\sigma^\pm$-polarised laser field respectively. 

\subsection{Transition lineshape}
Calculating the optical absorption of atom confined in NC is a challenging task and has only been discussed in the few works \cite{dutier.JOSAB.2003,zambon.OC.1997,vartanyan.PRA.1995}, and more recently by Peyrot \textit{et al.} \cite{peyrot.PRL.2019} who have pointed out nonlocality effects on electric susceptibility. Indeed, conventional dispersion theory is not sufficient the reproduce spectra recorded from NCs with a high fidelity. For our calculations, we follow the approach of Dutier \textit{et al.} \cite{dutier.JOSAB.2003}, who have shown that the thin cell transmission is an interferometric combination of responses associated with bulk reflection $I_{SR}^{lin}$ and bulk transmission $I_T^{lin}$, which reads
\begin{equation}
I_T = \left[1+r^2\exp(2ik\ell)\right]I_T^{lin} -2r I_{SR}^{lin},
\end{equation}
where $r$ is the windows' field reflection coefficient, $k$ is the wave number and $\ell$ is the cell thickness. In the limit of dilute vapour, the resonant contribution to transmitted light intensity is
\begin{equation}
S_t \propto \Re\left[{I_T}\right]\frac{E_{i}}{|Q|^2},
\end{equation}
where $E_{i}$ is the incident light field amplitude and $Q=1-r^2\exp(2ik\ell)$ is a factor associated with the cavity; $\Re$ denotes the real part. As $S_t$ is brackground-free, one can simply express the resonant contribution to the absorption spectrum as $S_a(\nu) = -1\times S_t(\nu)$. Expressions of $I_T^{lin}$ and $I_{SR}^{lin}$ are found after spatial integration of the transient atomic response obtained by averaging optical coherences of the density matrix over a Maxwellian distribution of velocities, see \cite{dutier.JOSAB.2003,zambon.OC.1997}. Transient optical coherences are expressed by solving Liouville equation of motion for the reduced density matrix $\sigma$ in the (time) steady-state regime:
\begin{equation}
v\frac{\partial \sigma}{\partial z} = -\frac{i}{\hbar}\left[ \mathcal H_l, \sigma\right] - \frac{1}{2} \left\lbrace\mathcal{R},\sigma \right\rbrace + \Lambda,
\end{equation}
where $v$ is a velocity class, $\mathcal H_l$ is the atom-light field Hamiltonian, $\mathcal{R}$ is a matrix including homogeneous relaxation processes and $\Lambda$ accounts for re-population (e.g. re-population of the ground state due to spontaneous decay of the excited state).
\begin{figure}
\centering
\includegraphics[width=9cm]{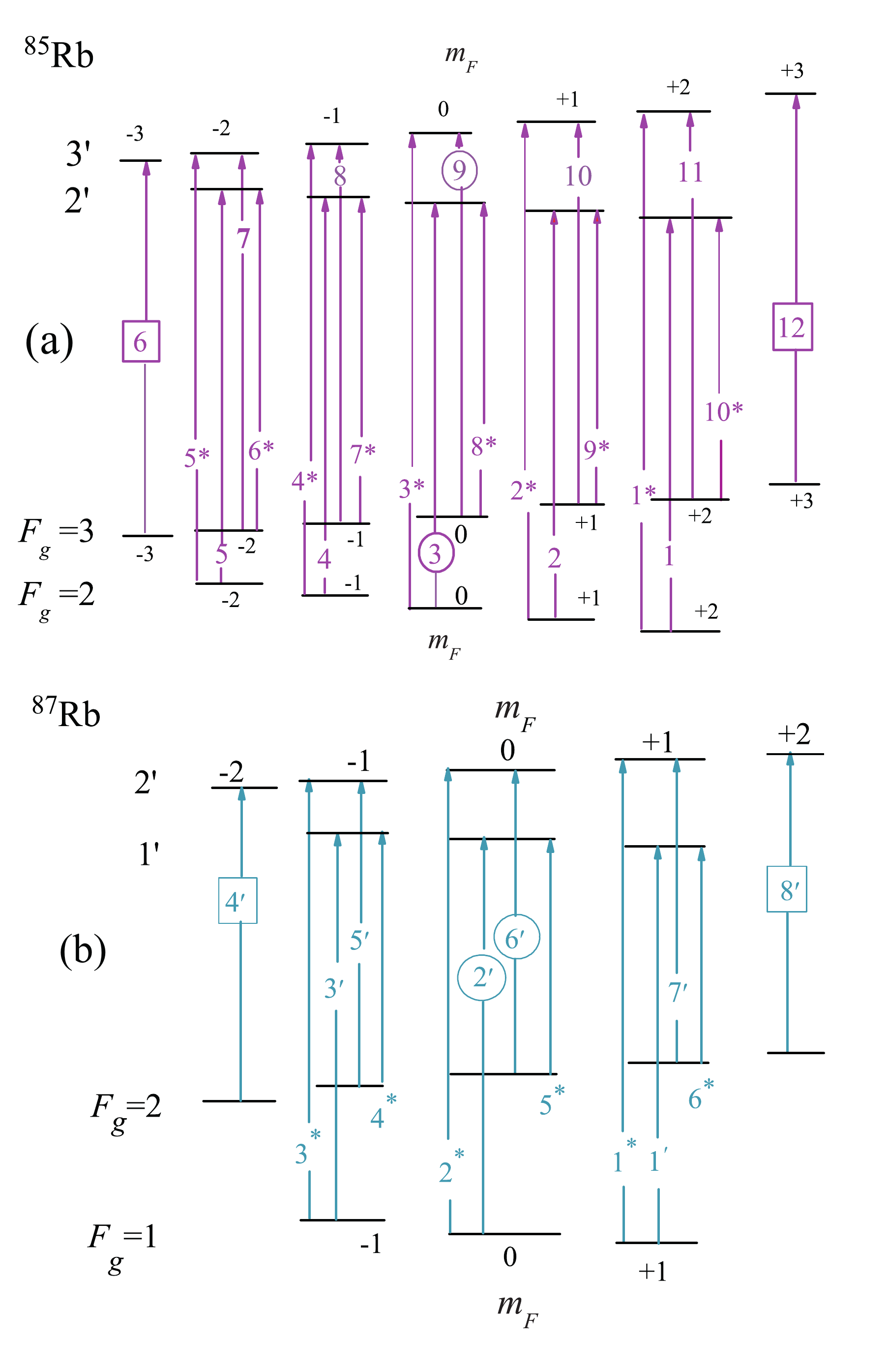}
\caption{Energy levels diagram of the D$_1$ line of (a) $^{85}$Rb and (b) $^{87}$Rb in the coupled basis $|F,m_F\rangle$, and atomic transitions for a $\pi$-polarised excitation ($\Delta m_F = 0$). Transitions labelled with a star disappear at high magnetic fields, labels in square indicate GT and labels in circle indicate MI transitions, see the text; the total number of transition is thirty six.}
\label{fig2}
\end{figure}

\begin{figure}
\centering
\includegraphics[width=8cm]{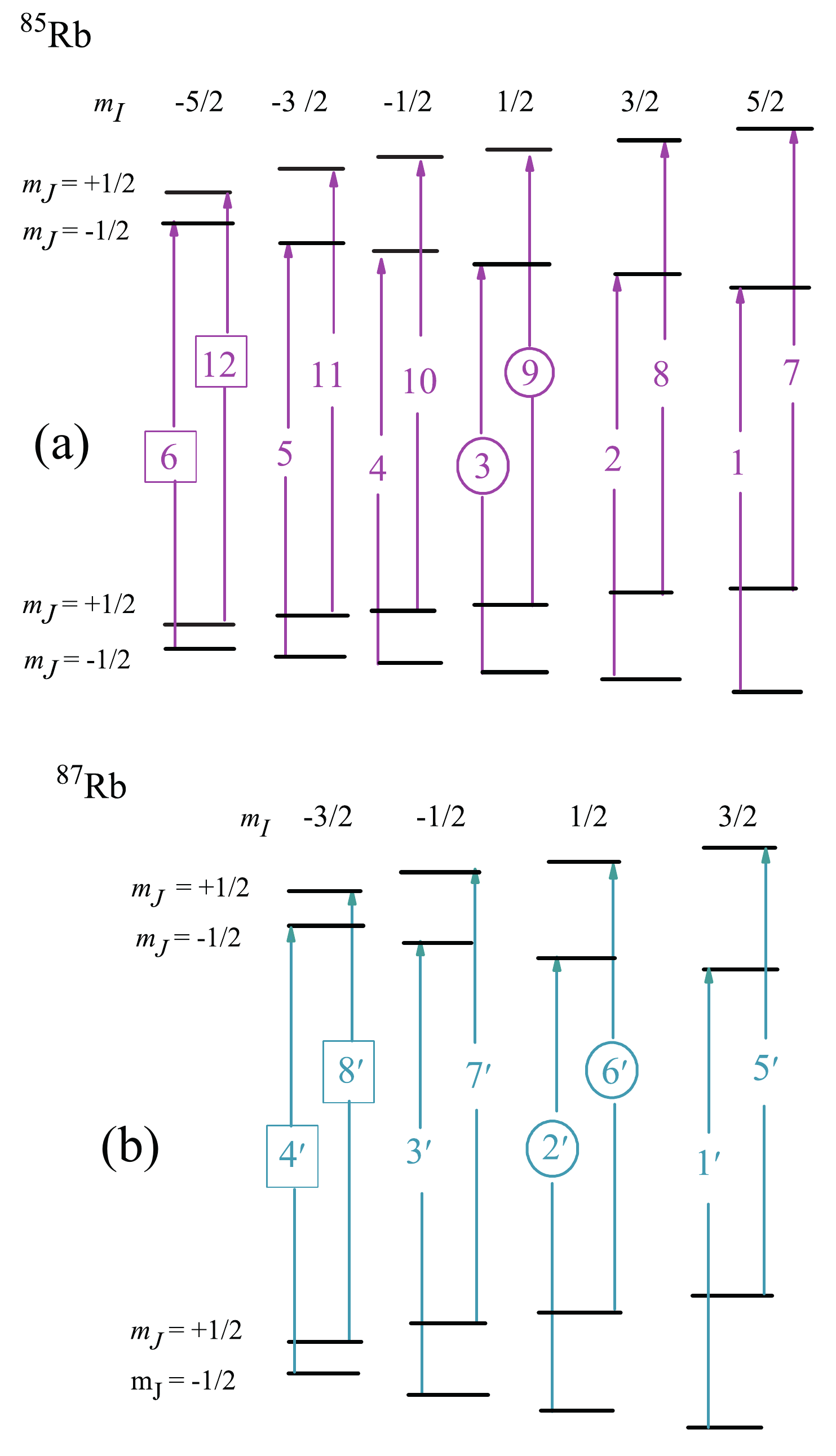}
\caption{Energy levels diagram of the D$_1$ line of (a) $^{85}$Rb and (b) $^{87}$Rb in the uncoupled basis $|m_I,m_J\rangle$, and atomic transitions for a $\pi$-polarised excitation ($\Delta m_I = \Delta m_J =0$).  Labels in square indicate GT and labels in circle indicate MI transitions, see the text; the total number of transition is twenty.}
\label{fig3}
\end{figure}

\section{Results and discussion}
\label{sec:results}

The behaviour of atomic transitions in a magnetic field is known to be different depending how strong is the external magnetic field. A simple way to understand if the applied magnetic field is weak or strong is to compare it with the caracteristic field $B_0= A_{hfs}/ \mu_B$, where $A_{hfs}$ is the hyperfine coupling constant of the ground state. For $^{85}$Rb one has $B_0 \sim 0.7$~kG while $B_0\sim 2.4$~kG for $^{87}$Rb; hence natural isotopes of Rb interact differently with the external field. When $B < B_0$, the splitting of the levels is better described by the total atomic angular momentum $\mathbf F = \mathbf J + \mathbf I$, where $\mathbf J$ is the total electronic angular momentum and $\mathbf I$ is the nuclear spin momentum, and its projection $m_F$; \textit{i.e.} in the coupled basis. Figure~\ref{fig2} shows the Rb D$_1$ line transitions in the coupled basis for a $\pi$-polarised laser excitation. As it is seen, there is a total of thirty six atomic transitions in this case. By increasing the magnetic field over $B_0$, one slowly decouples $\mathbf J$ and $\mathbf I$; hence $F$ is no longer a good quantum number and the splitting of the atomic levels is better described in the uncoupled basis $| m_J, m_I\rangle$, where $m_J$ and $m_I$ are the projections of the electronic angular momentum and the nuclear spin momentum respectively. This regime is often referred to as the hyperfine Paschen-Back (HPB) regime \cite{olsen.PRA.2011,sargsyan.OL.2012,weller.JPB.2012,sargsyan.JPB.2018}. In the HPB regime, only twenty atomic transitions should remain in the spectra as is seen in Fig.~\ref{fig3}. In addition to the reduced number of transitions, one also expects to have a lager frequency separation between transition than that for the case of $B < B_0$.

Atomic transitions $F_g=2, 3\rightarrow F_e=2', 3'$ of $^{85}$Rb D$_1$ line (a) and $F_g= 1, 2 \rightarrow F_e =1', 2'$ of $^{87}$Rb D$_1$ line (b), where primes indicate excited states quantum numbers, in the case of $\pi$-polarised laser excitation and for relatively small magnetic field are depicted in Fig.~\ref{fig2}. At zero magnetic field, the atomic transitions $|F_g,0\rangle \rightarrow |F_e = F_g,0'\rangle$, that is transitions \textcircled{3} and \textcircled{9} ($^{85}$Rb) as well as \textcircled{2}' and \textcircled{6}' ($^{87}$Rb), have zero probabilities due to selection rules \cite{sargsyan.JPB.2018}. However a giant increase of their probability in the presence of an external magnetic field has been reported in \cite{sargsyan.LPL.2014}; for this reason, theses transitions are often called magnetically-induced (MI) transitions. All others transitions presented in Fig.~\ref{fig2} verifying $\Delta F = F_g - F_e = 0,~\pm 1$ and $\Delta m_F = 0$ are allowed. Notheworthy, these MI transitions are very different from the MI transitions $\Delta F= F_e-F_g=\pm2$ studied in \cite{klinger.EPJD.2017,sargsyan.LPL.2014,sargsyan.OL.2019b}: the probabilities of the MI transitions studied in this paper tend to the probabilities of $^{85}$Rb and $^{87}$Rb guiding atomic transitions when $B \gg B_0$. Meanwhile the probabilities of MI transitions $\Delta F= F_e-F_g=\pm2$ for $B \gg B_0$ tend to zero. Theoretical calculations have confirmed this statement, see e.g. \cite{sargsyan.JPB.2018,sargsyan.LPL.2014,hakhumyan.EPJD.2012}. Equivalent diagrams of Fig.~\ref{fig2} in the uncoupled basis, that is when $B > B_0$, is presented in Fig.~\ref{fig3}. In both Fig.~\ref{fig2} and \ref{fig3} transitions whose labels are in square are the so-called guiding transitions (GT). An important feature of theses transitions is the following: the dependence of frequency shifts as well as transition probabilities has been observed to remain constant over the entire range $0.001 - 10$~kG \cite{sargsyan.JETPL.2015}. Transitions marked with stars disappear from the spectra at high magnetic field.

\subsection{Rb  D$_1$ line $\pi$-transitions spectrum under magnetic field}

\begin{figure}
\centering
\includegraphics[scale=1]{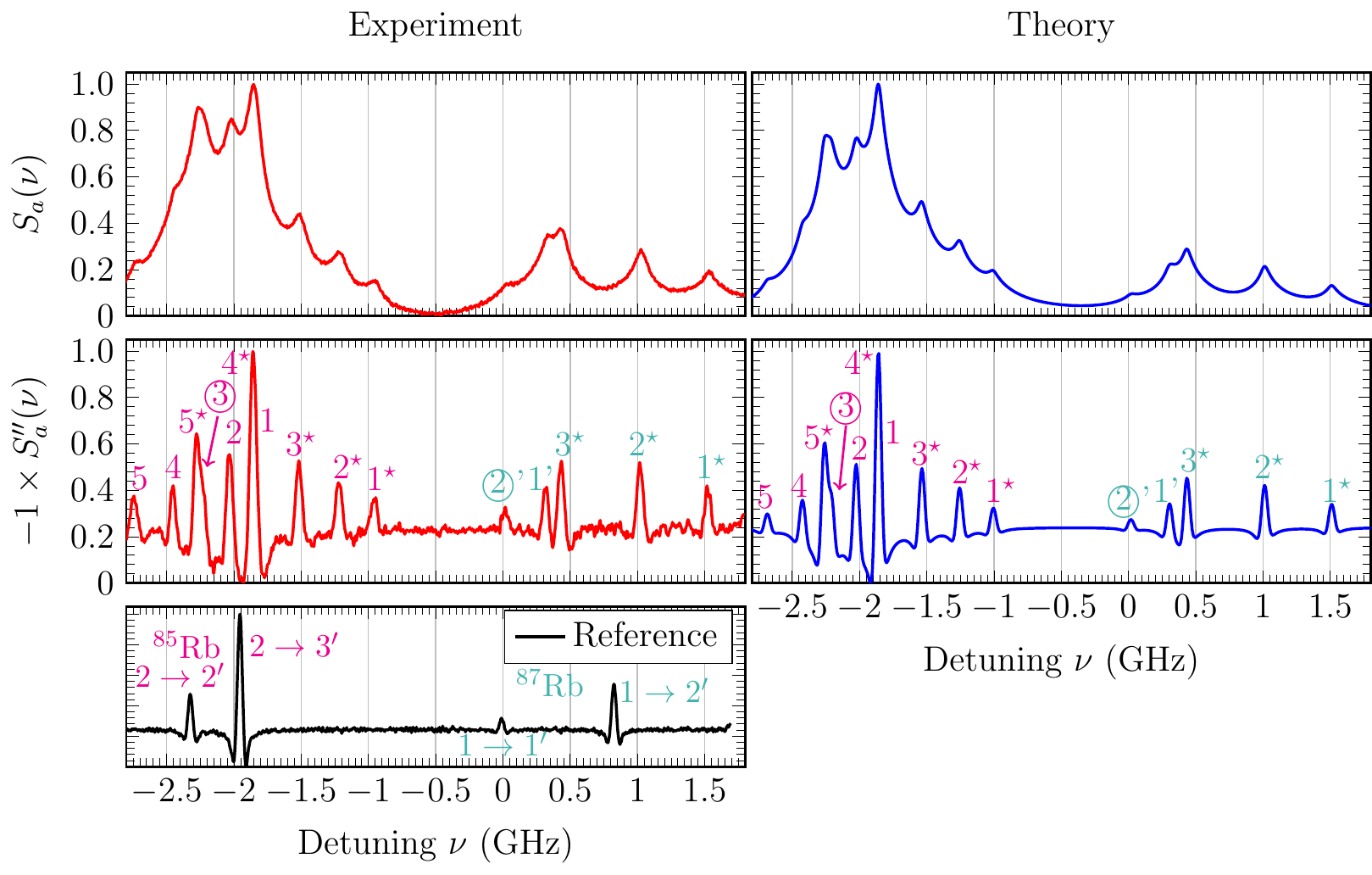}
\caption{High frequency range Rb D$_1$ line absorption and second derivative spectra recorded at $B =0.62$~kG for the thickness $\ell=\lambda/2$. Left panels -- experiment; right panels -- theory. The bottom left panel shows the reference Rb D$_1$ line spectrum recorded at zero magnetic field, where transition $1\rightarrow 1'$ of $^{87}$Rb was chosen as the zero frequency detuning. Both experimental and theoretical spectra have been normalized to each of their strongest transition. Atomic transitions of $^{87}$Rb are marked with primes, transitions disappearing at large magnetic fields are marked by stars; transitions labelled \textcircled{2}' and \textcircled{3} are magnetically-induced transitions, see the text.}
\label{fig4}
\end{figure}

In Figure~\ref{fig4}, left panels are experimental results while right panels are theoretical calculations. The top panels show the absorption spectrum of a $\pi$-polarised radiation tuned to the high frequency range of Rb D$_1$ line recorded from a NC with $\ell=\lambda/2= 398$~nm, at $B =0.62$~kG. For this graph, the laser power was about 40~$\mu$W and the beam diameter about 1 mm. The middle panels are the SD spectra directly derived from the absorption ones. The lower left panel is the frequency reference spectrum, where the atomic transition  $1 \rightarrow 1'$ of $^{87}$Rb was set as the zero frequency detuning: $\nu=\nu_l - \nu_{1\rightarrow1'}$, where $\nu_l$ is the laser frequency. Although separated by relatively small frequency intervals, atomic transitions are well resolved in the SD spectra; only transitions $5^\star$ and \textcircled{3} are -- partially -- overlapped. Despite the fact that small peaks are observable in the experimental absorption spectrum, fitting it with a specific lineshape to extract frequency positions and amplitudes is non-trivial since the line profile is neither Gaussian nor Lorentzian in this case, but rather close to the time-of-flight broadening profile \cite{meschede.Book.2007}. Meanwhile the SD technique gives correct frequency positions and amplitudes of atomic transition. In addition, MI transitions  \textcircled{2}' and \textcircled{3} have well detectable amplitudes at $B =0.62$~kG. A good agreement between experiment and theory is obviously seen from Fig.~\ref{fig4}.

\begin{figure}
\centering
\includegraphics[scale=1]{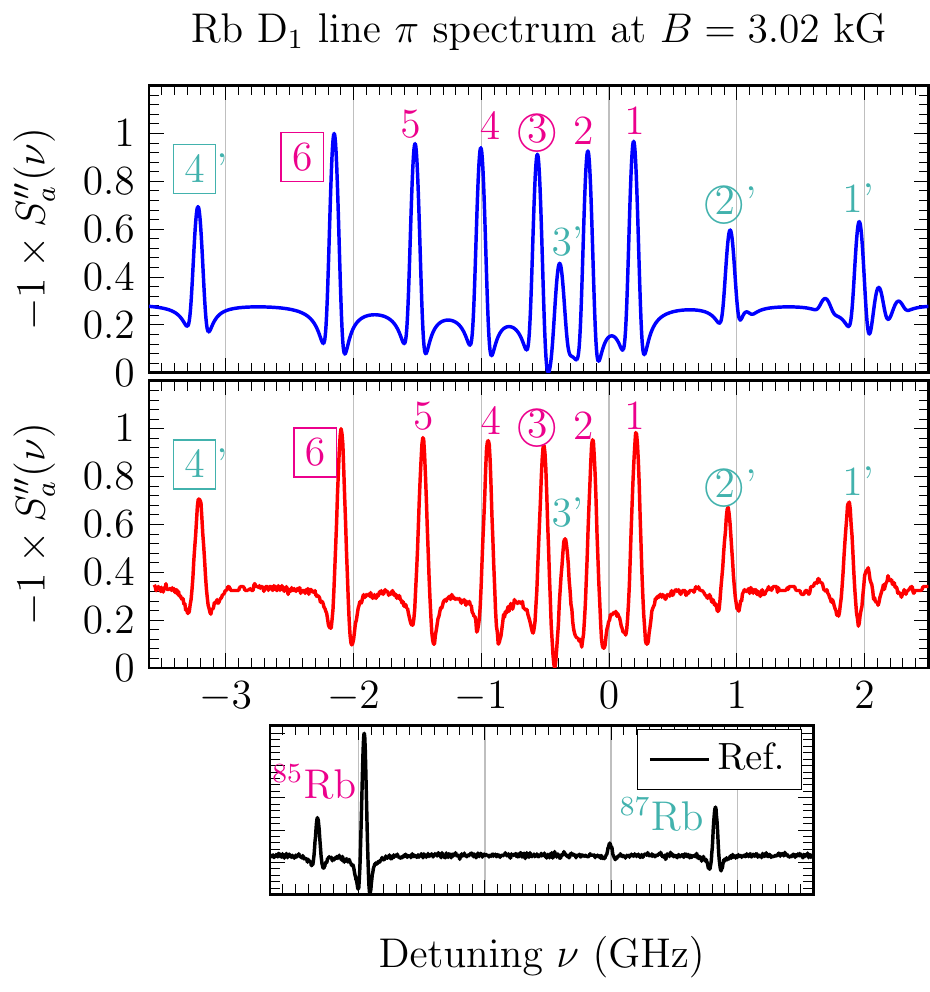}
\caption{High frequency range Rb D$_1$ line second derivative spectrum recorded at $B =3.02$~kG for the thickness $\ell=\lambda/2$. Top panel -- theory; middle panel -- experiment. The bottom left panel shows the reference Rb D$_1$ line spectrum recorded at zero magnetic field. Both experimental and theoretical spectra have been normalized to each of their strongest transition. Atomic transitions of $^{87}$Rb are marked with primes, transitions labelled  \textcircled{2}' and \textcircled{3} are magnetically induced transitions; transitions labelled $\fbox{6}$ and $\fbox{4}$' are guiding transitions, see the text.}
\label{fig5}
\end{figure}

Figure~\ref{fig5} shows the theoretical (top panel) and experimental (middle panel) SD $\pi$-spectrum of the high frequency range of Rb D$_1$ line, recorded from a NC with $\ell=\lambda/2$ at $B =3.02$~kG. As it is seen, all atomic transitions are completely resolved in the SD spectra. Moreover, one can see that the amplitude of transitions  1' to $\fbox{4}$' of $^{87}$Rb, including the magnetically-induced transition \textcircled{2}', are almost equal to each other. To be more precise, their amplitudes tend to that of the guiding transition of the same group, that is $\fbox{4}$', whose amplitude is constant. One can also observe the same behaviour for  transitions 1 to $\fbox{6}$ of $^{85}$Rb, including the magnetically-induced transition \textcircled{3}, whose amplitude is nearly the same as that of the GT $\fbox{6}$. Comparing the frequency intervals between $^{85}$Rb transitions, one notices that transitions 1 to $\fbox{6}$ are nearly equidistant. However, one may not say the same about $^{87}$Rb atomic transitions. To illustrate that, let us calculate the largest ratio of frequency intervals between consecutive transitions: $\Delta_{ij}/\Delta_{lm}=(\nu_i-\nu_j)/(\nu_l-\nu_m)$, where $i,j,l,m$ are transition labels and $\nu_{i,j,l,m}$ their respective frequencies. One gets $\Delta\textsubscript{\mathbox{6}5}/\Delta_{12}=1.75$ for $^{85}$Rb, while the largest ratio is $\Delta\textsubscript{3'\mathbox{4}'} /\Delta\textsubscript{1'\circled{2}'}=3$ for $^{87}$Rb. For $^{85}$Rb, these results are an evidence of an ongoing establishment of the HPB regime, also because $B> B_0(^{85}$Rb$)\approx 0.70$~kG; in contrast to $^{87}$Rb for which $B \sim B_0 (^{87}$Rb$)\approx 2.40$~kG. 

\begin{figure}
\centering
\includegraphics[scale=1]{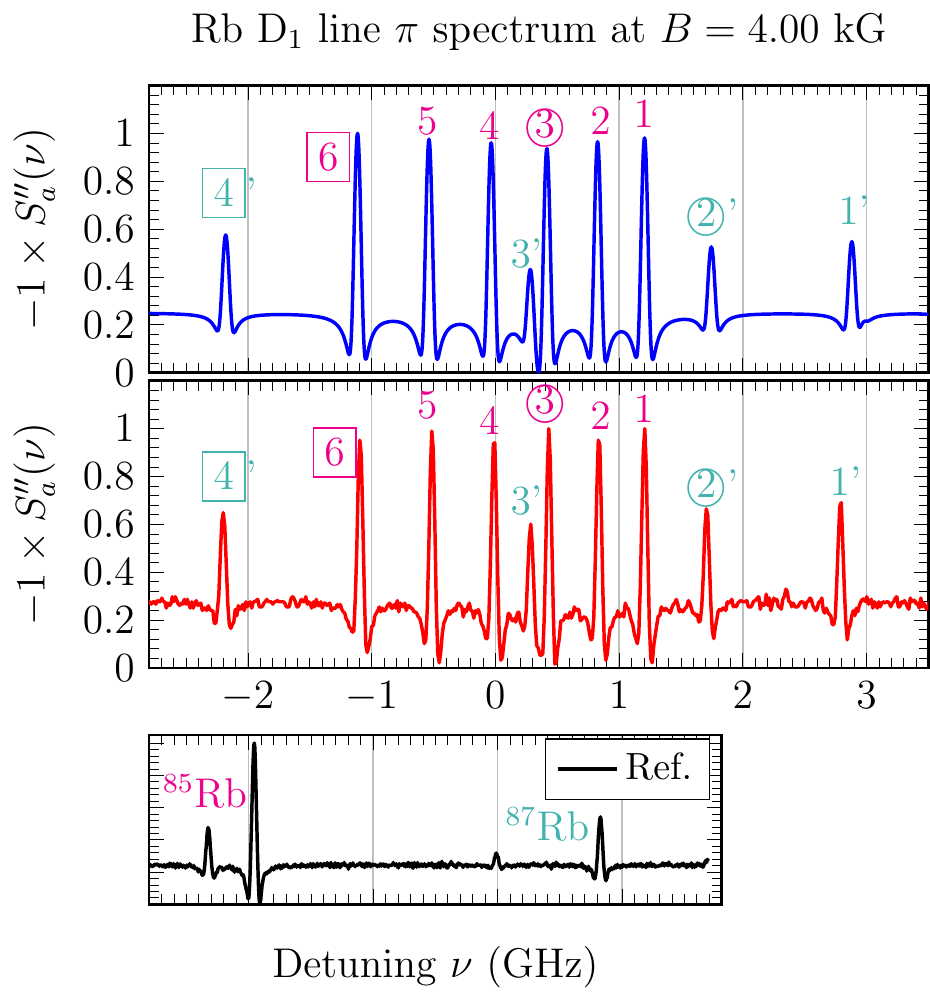}
\caption{High frequency range Rb D$_1$ line second derivative spectrum recorded at $B =4.00$~kG for the thickness $\ell=\lambda/2$. Top panel -- theory; middle panel -- experiment. The bottom left panel shows the reference Rb D$_1$ line spectrum recorded at zero magnetic field. Both experimental and theoretical spectra have been normalized to each of their strongest transition. Atomic transitions of $^{87}$Rb are marked with primes, transitions labelled  \textcircled{2}' and \textcircled{3} are magnetically induced transitions; transitions labelled $\fbox{6}$ and $\fbox{4}$' are guiding transitions, see the text.}
\label{fig6}
\end{figure}

Figure~\ref{fig6} shows the theoretical (top panel) and experimental (middle panel) SD $\pi$-spectrum of the high frequency range of Rb D$_1$ line, recorded from a NC with $\ell=\lambda/2$ at $B = 4$~kG. Once again, on can see that all atomic transitions are completely resolved. This time, the largest frequency interval ratio for $^{85}$Rb is now $\Delta\textsubscript{\mathbox{6}5}/\Delta_{12}=1. 5$, while one gets $\Delta\textsubscript{3'\mathbox{4}'} /\Delta\textsubscript{1'\circled{2}'}=2.2$ for $^{87}$Rb. It should be noted that for $B\gg B_0 (^{87}$Rb$)$, and consequently $B\gg B_0 (^{85}$Rb$)$, the above mentioned ratios for both $^{85}$Rb and $^{87}$Rb tends to a value close to unity with increasing magnetic field. Nevertheless, the ratio does not exactly reach 1 because of different values of nuclear spin projection $m_I$ for these atomic transitions, see Fig.~\ref{fig3}. As for the  transition amplitudes, one can observe that the amplitude of transitions 1 to 5 approaches that of the GT $\fbox{6}$, while the amplitude of transition 1' to 3' approaches that of $\fbox{4}$'; which is an additional confirmation of the onset of the Paschen-Back regime.

\begin{figure}
\centering
\includegraphics[scale=1]{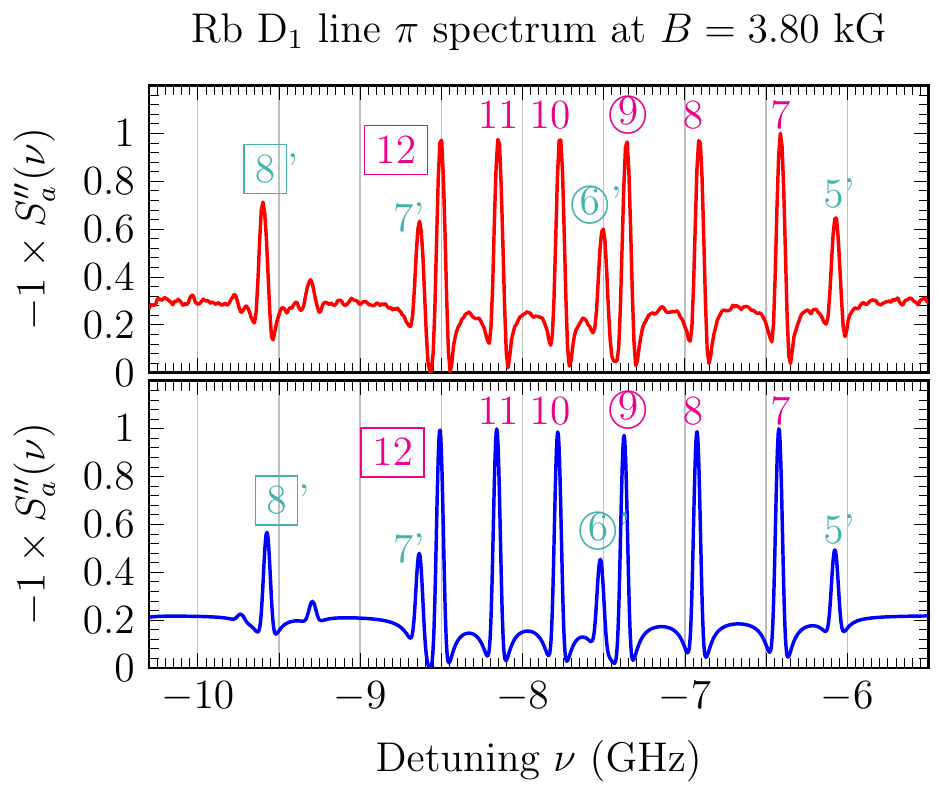}
\caption{Low frequency range Rb D$_1$ line second derivative spectrum recorded at $B =3.80$~kG for the thickness $\ell=\lambda/2$. Top panel -- theory; middle panel -- experiment. The zero frequency detuning is set on the transition $1\rightarrow 1'$ of $^{87}$Rb.}
\label{fig7}
\end{figure}

As noted here-before, one expects to record, in the HPB regime, a total of twenty atomic transitions for both $^{85}$Rb and $^{87}$Rb respectively. However, only ten can be seen in Fig.~\ref{fig6}. Theoretical calculations show that all twenty transitions span over approximately 13 GHz at $B = 4.00$~kG. However, the region of linear scanning of laser frequency does not exceed 5~GHz for the laser used in our experiments. Therefore, we had to separately record the high and low frequency regions of Rb D$_1$ line spectrum in a magnetic field. Figure~\ref{fig7} thus shows the theoretical (top panel) and experimental (middle panel) SD $\pi$-spectrum of the low frequency range of Rb D$_1$ line, recorded from a NC with $\ell=\lambda/2$ at $B = 3.80$~kG. In the graphs, atomic transitions of $^{87}$Rb are marked with primes; transitions \textcircled{6}' and \textcircled{9} are MI transitions while transitions $\fbox{8}$' and $\fbox{12}$ are GT. As is seen, there are four atomic transitions of $^{87}$Rb and six for $^{85}$Rb, which bring the total number of transitions for both Figs.~\ref{fig6} and \ref{fig7} to twenty as was expected.

\subsection{Dark resonance formation on MI transitions}

\begin{figure}
\centering
\includegraphics[scale=1]{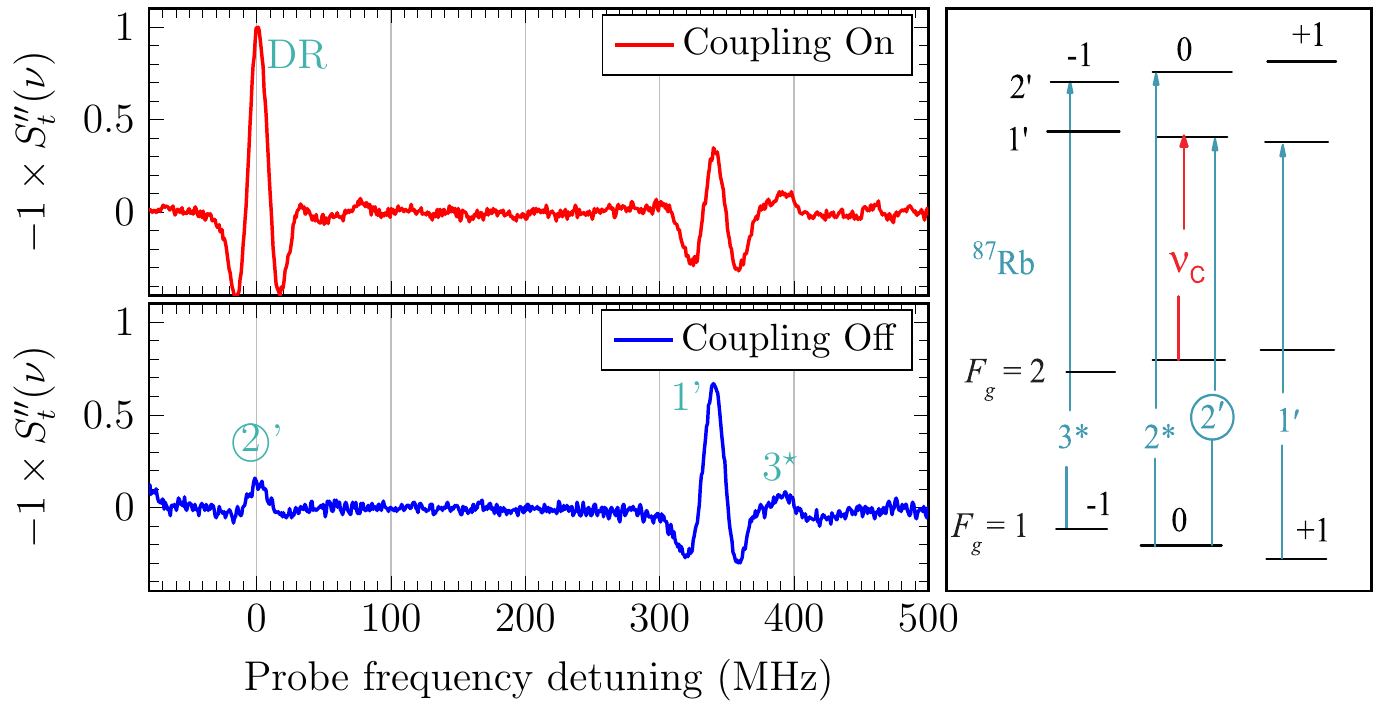}
\caption{Experimental SD transmission spectrum recorded at $B=0.70$~kG when the coupling laser tuned to the transition $|2,0\rangle \rightarrow |1',0'\rangle$ is On -- top panel, and Off -- bottom panel. Both curves have been normalised to the DR amplitude. Zero probe frequency detuning corresponds to transition frequency of the MI transition \textcircled{2}' at $B=0.70$~kG. The right inset shows the diagram of the pump-probe configuration.}
\label{fig8}
\end{figure}

Now, let us to demonstrate that using an additional (coupling) laser, it is possible to form dark resonances (DRs) on MI transitions via electromagnetically induced transparency (EIT). Experimental studies were carried out on a setup similar to that described in \cite{sargsyan.OL.2019b}. Radiations of two continuous-wave narrow-band ($<1~$MHz linewidth) ECDL, tunable around $\lambda = 795$~nm, were used to form a pump-probe configuration. The coupling beam frequency $\nu_c$ was fixed on the transition $|2,0\rangle \rightarrow |1',0'\rangle$ (also labeled $5^\star$) of $^{87}$Rb, while the probe beam frequency $\nu_p$ was scanned around transitions $1\rightarrow 1',2'$, covering the MI transition \textcircled{2}'. The energy levels and transitions for the $\Lambda$-system of $^{87}$Rb D$_1$ line relevant for the formation of DR are shown in the inset of Fig.~\ref{fig8}. The $1$~mm-large coupling and probe beams were combined in a NC with a thickness $\ell=2\lambda \sim 1.5~\mu$m. As shown in \cite{sargsyan.OL.2019b}, with such a small thickness, the formation of DR resonance is still possible. The temperature of the NC's reservoir was kept at $110~^\circ$C, corresponding to a vapour number density of $\mathcal N \sim 10^{13}$~cm$^{-3}$. The employed experimental configuration allowed to individually control the power of both beams, $P_c =30$~mW for the coupling and $P_p =0.4$~mW for the probe, and their polarisations. Two polarising beam splitters were used to purify initial linear polarisations of both the coupling and probe laser beams. Since the coupling and the probe lasers have collinear polarisations, oriented along the magnetic field, the two beams are directed at an angle $\varphi \sim 20$~mrad so that only the probe beam falls onto the photodetector. The configuration of the magnetic measurement is similar to that presented in the lower left inset of Fig.~\ref{fig1}. 

In Fig.~\ref{fig8}, the top panel shows the experimental SD transmission spectrum of the probe radiation tuned around $1\rightarrow1',2'$ transitions, at $B =0.70$~kG, when the coupling laser is on. The bottom panel shows the same spectrum recorded when the coupling laser is turned off. As is seen in the top panel, a dark resonance \cite{fleischhauer.RMP.2005} is formed via EIT when the probe is resonant with transition \textcircled{2}'. The DR resonance demonstrates about five times times greater transmission than the MI transition \textcircled{2}' when the coupling laser is absent. In addition to the DR formation, the amplitude of the transition $1'$ demonstrates an increase in absorption because of the re-pumping effect caused by the coupling laser: the coupling laser transfers some population from the state $F_g=2$ to $F_g=1$ through the excited state $F_e=1$, which leads to an increase of absorption of the state $F_g=1$. Note that a much narrower spectral width of DR can be attained by using cm-long cell with buffer gas, and employing coherently coupled probe and coupling radiations derived from a single laser beam source \cite{fleischhauer.RMP.2005}.

\section{Conclusion}
\label{sec:conc}
We have demonstrated that despite a large Doppler width of atomic transitions of rubidium, the SD technique performed on the absorption spectra of Rb atomic vapours confined in a NC with the thickness $\ell= \lambda/2=398$~nm allows to realise close to Doppler-free spectroscopy. Narrow linewidth and linearity of the signal response with respect to transition probabilities allow us to resolve, in an external transverse magnetic field, a large number of atomic transitions in the natural Rb vapour. In particular, we have shown that the probabilities of MI transitions $|1,0\rangle \rightarrow|1',0'\rangle$ and $|2,0\rangle \rightarrow |2',0'\rangle$ of $^{87}$Rb and, $|2,0\rangle \rightarrow |2',0'\rangle$ and $|3,0\rangle \rightarrow |3',0'\rangle$ of $^{85}$Rb, which are null at zero magnetic field, show a gigantic increase in the probability with an increase in the magnetic field. Theoretical calculations are in a very good agreement with experimental results.   

The excellent spectral resolution of atomic transitions obtained with the SD spectra also allowed us to quantitatively trace the onset of the hyperfine Paschen-Back regime and its further development to the full HPB regime with increasing transverse magnetic field. This was achieved by measuring the frequency intervals between atomic transitions and comparing their amplitudes. We have also verified experimentally that the total number of the detected atomic transition of Rb D$_1$ line in the HPB regime is twenty: eight belong to $^{87}$Rb and twelve to $^{85}$Rb. Meanwhile, when $B \ll B_0$, the total number of atomic transitions is thirty six: fourteen for $^{87}$Rb and twenty two for $^{85}$Rb.

 Finally, we have demonstrated the possibility to form a dark resonance via EIT on a magnetically induced transition of $^{87}$Rb using a coupling and a probe laser, when a transverse magnetic field is applied on the vapour. We have measured a five-fold increase in the transmission of the probe radiation when the coupling laser was on.
 	
 	 We should note that the recent development of a glass NC \cite{peyrot.OL.2019}, much easier to manufacture than sapphire-made NC used in the present work, could make NCs and the SD technique available for a wider range of researchers and engineers.

\ack The authors acknowledge A Tonoyan, A Papoyan and C Leroy for fruitful discussions. This work was supported by the Science Committee of the Ministry of Education, Science, Culture and Sport of the Republic of Armenia (projects no. SCS 18T-1C018  and no. SCS 19YR-1C017).


\end{document}